\documentclass[twocolumn,tighten]{aastex63}

\usepackage{amsmath}
\usepackage{xcolor}
\usepackage{wasysym}
\usepackage{enumitem}
\usepackage{comment}
\usepackage{verbatim}

\definecolor{my_color}{HTML}{3a18b1}

\definecolor{new_color}{HTML}{CF0000}% this is a maroon
\definecolor{new_black}{HTML}{000000}% this is a maroon
\usepackage{xfrac}

\newcommand\bedit[1]{\textcolor{new_black}}

\newcommand{\be}{\begin{equation}}
\newcommand{\ee}{\end{equation}}

\submitjournal{AJ}

\shorttitle{Beta Pic Exomoon Limits}
\shortauthors{Macias et al.}
\begin{document}

%Beta Pic params WITH NO MOON
\newcommand{\smaonenomoon}{10.03}
\newcommand{\usmaonenomoon}{\ensuremath{^{+0.11}_{-0.10}}}
\newcommand{\ecconenomoon}{0.1094}
\newcommand{\uecconenomoon}{\ensuremath{^{+0.0077}_{-0.0077}}}
\newcommand{\inconenomoon}{88.991}
\newcommand{\uinconenomoon}{\ensuremath{^{+0.014}_{-0.014}}}
\newcommand{\aoponenomoon}{212.5}
\newcommand{\uaoponenomoon}{\ensuremath{^{+7.2}_{-6.4}}}
\newcommand{\panonenomoon}{31.773}
\newcommand{\upanonenomoon}{\ensuremath{^{+0.015}_{-0.015}}}
\newcommand{\tauonenomoon}{0.774}
\newcommand{\utauonenomoon}{\ensuremath{^{+0.018}_{-0.017}}}
\newcommand{\smatwonomoon}{2.688}
\newcommand{\usmatwonomoon}{\ensuremath{^{+0.083}_{-0.086}}}
\newcommand{\ecctwonomoon}{0.235}
\newcommand{\uecctwonomoon}{\ensuremath{^{+0.012}_{-0.012}}}
\newcommand{\inctwonomoon}{88.94}
\newcommand{\uinctwonomoon}{\ensuremath{^{+0.14}_{-0.14}}}
\newcommand{\aoptwonomoon}{61}
\newcommand{\uaoptwonomoon}{\ensuremath{^{+8.5}_{-11}}}
\newcommand{\pantwonomoon}{31.18}
\newcommand{\upantwonomoon}{\ensuremath{^{+0.068}_{-0.057}}}
\newcommand{\tautwonomoon}{0.825}
\newcommand{\utautwonomoon}{\ensuremath{^{+0.020}_{-0.026}}}

\newcommand{\plxnomoon}{51.39}
\newcommand{\uplxnomoon}{\ensuremath{^{+0.11}_{-0.12}}}
\newcommand{\mbnomoon}{10.02}
\newcommand{\umbnomoon}{\ensuremath{^{+0.29}_{-0.29}}}
\newcommand{\mcnomoon}{9.030}
\newcommand{\umcnomoon}{\ensuremath{^{+0.080}_{-0.079}}}
\newcommand{\mstarnomoon}{1.701}
\newcommand{\umstarnomoon}{\ensuremath{^{+0.022}_{-0.021}}}
\newcommand{\raoffsetspherebnomoon}{-0.10}
\newcommand{\uraoffsetspherebnomoon}{\ensuremath{^{+0.33}_{-0.33}}}
\newcommand{\decoffsetspherebnomoon}{0.06}
\newcommand{\udecoffsetspherebnomoon}{\ensuremath{^{+0.35}_{-0.35}}}
\newcommand{\errscalegravitynomoon}{1.31}
\newcommand{\uerrscalegravitynomoon}{\ensuremath{^{+0.36}_{-0.21}}}

%Beta Pic params WITH MOON
\newcommand{\smaone}{10.02}
\newcommand{\usmaone}{\ensuremath{^{+0.11}_{-0.10}}}
\newcommand{\eccone}{0.1072}
\newcommand{\ueccone}{\ensuremath{^{+0.0086}_{-0.0091}}}
\newcommand{\incone}{88.995}
\newcommand{\uincone}{\ensuremath{^{+0.016}_{-0.015}}}
\newcommand{\aopone}{211.1}
\newcommand{\uaopone}{\ensuremath{^{+7.3}_{-6.6}}}
\newcommand{\panone}{31.771}
\newcommand{\upanone}{\ensuremath{^{+0.017}_{-0.017}}}
\newcommand{\tauone}{0.769}
\newcommand{\utauone}{\ensuremath{^{+0.019}_{-0.018}}}
\newcommand{\smatwo}{2.66}
\newcommand{\usmatwo}{\ensuremath{^{+0.11}_{-0.12}}}
\newcommand{\ecctwo}{0.234}
\newcommand{\uecctwo}{\ensuremath{^{+0.012}_{-0.012}}}
\newcommand{\inctwo}{88.92}
\newcommand{\uinctwo}{\ensuremath{^{+0.14}_{-0.15}}}
\newcommand{\aoptwo}{57}
\newcommand{\uaoptwo}{\ensuremath{^{+12}_{-89}}}
\newcommand{\pantwo}{31.21}
\newcommand{\upantwo}{\ensuremath{^{+0.44}_{-0.074}}}
\newcommand{\tautwo}{0.82}
\newcommand{\utautwo}{\ensuremath{^{+0.028}_{-0.27}}}
\newcommand{\periodmoonone}{414}
\newcommand{\uperiodmoonone}{\ensuremath{^{+634}_{-380}}}
\newcommand{\incmoonone}{89}
\newcommand{\uincmoonone}{\ensuremath{^{+39}_{-39}}}
\newcommand{\panmoonone}{178}
\newcommand{\upanmoonone}{\ensuremath{^{+125}_{-117}}}
\newcommand{\taumoonone}{0.53}
\newcommand{\utaumoonone}{\ensuremath{^{+0.31}_{-0.35}}}
\newcommand{\plx}{51.40}
\newcommand{\uplx}{\ensuremath{^{+0.12}_{-0.11}}}
\newcommand{\mmoonone}{0.04}
\newcommand{\ummoonone}{\ensuremath{^{+0.11}_{-0.030}}}
\newcommand{\mb}{10.02}
\newcommand{\umb}{\ensuremath{^{+0.29}_{-0.29}}}
\newcommand{\mc}{9.030}
\newcommand{\umc}{\ensuremath{^{+0.079}_{-0.079}}}
\newcommand{\mstar}{1.710}
\newcommand{\umstar}{\ensuremath{^{+0.025}_{-0.024}}}
\newcommand{\raoffsetsphereb}{-0.09}
\newcommand{\uraoffsetsphereb}{\ensuremath{^{+0.33}_{-0.33}}}
\newcommand{\decoffsetsphereb}{0.07}
\newcommand{\udecoffsetsphereb}{\ensuremath{^{+0.35}_{-0.35}}}
\newcommand{\errscalegravity}{1.38}
\newcommand{\uerrscalegravity}{\ensuremath{^{+0.40}_{-0.24}}}

%%%EXTRA BETA PIC COMMANDS

\newcommand{\betapic}{\ensuremath{\beta} Pic}
\newcommand{\betapictoris}{\ensuremath{\beta} Pictoris}

\newcommand{\betapicb}{\ensuremath{\beta} Pic b}
\newcommand{\betapicc}{\ensuremath{\beta} Pic c}

\newcommand{\betapictorisb}{\ensuremath{\beta} Pictoris b}
\newcommand{\betapictorisc}{\ensuremath{\beta} Pictoris c}

\title{First Astrometric Limits on Binary Planets and Exomoons orbiting $\beta$ Pictoris b}

\author[0009-0003-5946-3616]{Isabella Macias}
\affiliation{Department of Earth, Atmospheric, and Planetary Sciences, Massachusetts Institute of Technology, 77 Massachusetts Avenue, Cambridge, MA 02139, USA}
\affiliation{Department of Astronomy, University of Florida, Bryant Space Science Center, Stadium Road, Gainesville, FL 32611, USA}

\author[0000-0001-9827-1463]{Sydney A. Jenkins} 
\affiliation{Department of Physics and Kavli Institute for Astrophysics and Space Research, Massachusetts Institute of Technology, 77 Massachusetts Avenue, Cambridge, MA 02139, USA}
\altaffiliation{NSF Graduate Research Fellow}

\author[0000-0001-7246-5438]{Andrew Vanderburg}
\affiliation{Center for Astrophysics $\vert$ Harvard \& Smithsonian, 60 Garden Street, Cambridge, MA 02138, USA}
\affiliation{Department of Physics and Kavli Institute for Astrophysics and Space Research, Massachusetts Institute of Technology, 77 Massachusetts Avenue, Cambridge, MA 02139, USA}
\altaffiliation{Sloan Research Fellow}

%\author{Saul A. Rappaport}
%\affiliation{Department of Physics and Kavli Institute for Astrophysics and Space Research, Massachusetts Institute of Technology, 77 Massachusetts Avenue, %Cambridge, MA 02139, USA}

\correspondingauthor{Isabella Macias}
\email{maciasi@mit.edu}



\begin{abstract}

The search for exomoons, or moons in other star systems, has attracted significant interest in recent years, driven both by advancements in detection sensitivity and by the expanding population of known exoplanets. The \betapictoris\ system is a particularly favorable target, as its proximity and directly imaged planets allow for precise astrometric monitoring. We present astrometric constraints on the presence of binary planets and exomoons in the \betapictoris\ system using archival observations from the GRAVITY interferometer and SPHERE instruments. We calculate these limits by modeling the motion of the two orbiting planets and introducing an additional perturbation to the model that simulates the astrometric motion caused by an exomoon orbiting the planet \betapictorisb. We find that for short orbital periods ($\approx50$ days), a lunar companion is only allowed if its mass remains below $\approx 180~M_{\oplus}$ ($0.6~M_{\text{Jup}}$) at $3\sigma$ confidence. At intermediate periods near 300 days, we exclude moons more massive than $\approx 65~M_{\oplus}$ ($0.2~M_{\text{Jup}}$) at $3\sigma$ confidence. At longer orbital periods, we place the tightest constraints, ruling out any potential exomoon above $\approx 50~M_{\oplus}$ ($0.15~M_{\text{Jup}}$) at $700$ days and $\approx 30~M_{\oplus}$ ($0.1~M_{\text{Jup}}$) at $1,100$ days (both at $3\sigma$ confidence). These results place the first astrometric constraints on moons and binary planets in the Beta Pictoris system and demonstrate the sensitivity of interferometric observations for exomoon studies. 

\end{abstract}



\section{Introduction}\label{introduction}

As thousands of exoplanets have been discovered in the past 30 years, interest in whether moons orbit these worlds has grown. Despite the prevalence of moons in the Solar System (over 400 known\footnote{\url{https://ssd.jpl.nasa.gov/sats/discovery.html}}) and the detection of a few candidates outside the Solar System \citep[e.g.][]{Teachey_2018, Lazzoni_2020, Kipping_2022}, no extrasolar moons, or exomoons, have been confirmed. Nevertheless, given the well-established moon formation pathways benchmarked in the Solar System, theoretical studies suggest that exomoons should exist. One dominant mechanism in the Solar System is co-formation with the planets in circumplanetary disks, where moons form alongside their host planets in a gas-rich disk, similar to the Galilean moons of Jupiter \citep{canup&ward}. Another formation pathway is capture events, where a passing object becomes gravitationally bound to a planet. This process has been proposed as a key mechanism for the formation of moons on eccentric, inclined, and often retrograde orbits in our Solar System \citep{hamers&zwart}, including Triton \citep{Agnor_2006}. Furthermore, debris generated from massive impacts, such as the hypothesized origin of the Earth-Moon system, could also lead to exomoon formation \citep{bar&sayl}.

The hunt for exomoons is especially important for understanding the formation and evolution of planetary systems. Just as the Moon plays a key role in Earth's orbital stability and tides, exomoons are expected to gravitationally influence their host planets, potentially affecting their rotational dynamics and long-term climate stability \citep{laskar, Poon_2024}. The successful detection of an exomoon would also provide valuable constraints on lunar habitability and planetary migration theories, as the survival, orbital architecture, and tidal evolution of moons depend sensitively on how and when their host planets migrate \citep{barnes&obrien, sucerquia2019, sucerquia2020, martinez-rodriguez}.

Despite the importance of discovering exomoons, their detection remains extremely challenging. The signals from moons are often too subtle and faint to detect directly as a result of the limited resolution and sensitivity of current telescopes. Indirect methods, such as analyzing light curve variations during planetary transits, offer alternative ways to detect exomoons \citep{Kipping_2015, Teachey_2018, Kipping_2021}. Researchers search for signals such as transit timing variations \citep[TTVs,][]{kipping2009a, kipping2020a, kipping2020b}, transit duration variations \citep[TDVs,][]{Kipping_2009}, and transit radius variations \citep[TRVs,][]{Rodenbeck_2020}. %These changes can be caused by the gravitational influence of an orbiting moon, providing indirect evidence for its presence. 
From these methods, a handful of exomoon candidates have been proposed, including Kepler-1625b-i, a potential Neptune-sized moon orbiting a super-Jovian planet \citep{Teachey_2018}, and Kepler-1708b-i, a potential 2.6 Earth-radii moon orbiting a Jupiter-sized planet at a distance of 12 planetary radii \citep{Kipping_2022}. However, none of these candidates have been confirmed, and the detections are controversial \citep[e.g.][]{Kreidberg2019ApJL, Heller2024NatAs}. We note that similar methods can also be used to detect binary planets \citep{Chakraborty&Kipping_2022}, but none have been identified.

%Other detection methods that have been proposed include direct imaging \citep{Lazzoni_2020}, to more recent innovations involving planetary property analysis, such as linking obliquity to the presence of moons \citep{Poon_2024}. Additionally, deep learning models, particularly convolutional neural networks (CNNs), have been explored as a tool to scan light curves for exomoon signals, offering a promising approach for automated detection \citep{Teachey_2021}.  However, no exomoons have been confirmed, as all methods face challenges, including detection limits, observational constraints, and model uncertainties.
%kipping2009a, 
%Early theoretical work established that transit timing and duration variations could reveal moons \citep{}.

With no confirmed exomoons, research has instead focused on placing increasingly strong limits on their possible presence. For instance, a considerable amount of work has gone into placing upper bounds on the masses and radii of exomoons based on transit observations \citep[e.g.][]{kipping_2013, Kipping_2014, Kipping_2015, kipping2020a, kipping2020b}. More recently, other detection methods have been used to place constraints, including radial velocity (RV) observations of directly imaged planets \citep{Vanderburg2018AJ, vanderburg2021, Ruffio2023AJ, Horstman2024AJ}. These findings demonstrate that, although the confident detection of exomoons is still elusive, current methods can place constraints on their masses and orbits. %\citet{vanderburg2021} used RV measurements of HR 8799 to place constraints on the possible presence of exomoons and binary planets around HR 8799 b, c, and d.

Several studies have suggested that astrometry can be used to identify and constrain lunar companions within a system \citep{cabrera2007, agol2015, vanWoerkom2024A&A}. This is a particularly exciting method given the recent advancements in measuring precise astrometry of exoplanets using long-baseline interferometry. 
%Early work by \citet{cabrera2007} proposed detecting planet-moon systems through mutual events, and \citet{agol2015} introduced spectroastrometry as a way to distinguish planet and moon signals in direct imaging data \citep[see also][]{vanWoerkom2024A&A}. 
Recently, \citet{lacour2021} showcased the high-precision astrometric capabilities of the GRAVITY instrument \citep{gravity2017} on the Very Large Telescope Interferometer (VLTI), which can make astrometric measurements with precision of tens of microarcseconds. Building on these capabilities, \citet{winterhalder2025astrometricexomoondetectionmeans} demonstrated through simulations and VLTI/GRAVITY observations that moons as small as $\sim$0.14 $M_{Jup}$ could already be inferred around the nearby (d$=$26.9 pc) AF Lep b system, with future facilities expected to reach Earth-like moons in habitable zones. Complementary work by \citet{wagner2025astrometricmethodsdetectingexomoons} presented a simulation framework for detecting exomoons around imaged giant planets, showing that Earth-mass moons in the habitable zone of $\alpha$ Centauri A could be detected with future space- or ground-based facilities. Most recently, \citet{Kral2025arXiv} placed constraints on moons orbiting the brown dwarf HD 206893 B with GRAVITY astrometry and detected a tentative candidate. Together, these studies highlight how astrometry is emerging as a promising pathway toward the first confirmed detections of exomoons.

One of the most promising targets for astrometric studies is the Beta (\(\beta\)) Pictoris system, a young and nearby planetary system that has been extensively studied since the discovery of its circumstellar debris disk in the 1980s \citep{smith&terrile_1984}. This system hosts two directly imaged giant planets, \(\beta\) Pic b and c, making it ideal for testing astrometric detection techniques. \(\beta\) Pic b, located at a semi-major axis of \( \cong 9.8 \) AU, was discovered through direct imaging, while \(\beta\) Pic c, at \( \cong 2.7 \) AU, was \bedit{initially} discovered through RV measurements \bedit{and subsequently directly detected with interferometric astrometry \citep{lagrange2019a, lagrange2020, Nowak_2020}}. \citet{lagrange2020} combined astrometry and RV data to constrain the orbits and masses of both planets, finding that \(\beta\) Pic b has a mass of \( 10\text{--}11 \, M_{\text{Jup}} \) and \(\beta\) Pic c has a well-constrained mass of \( 7.8 \pm 0.4 \, M_{\text{Jup}} \).

In this work, we use archival data from \citet{lagrange2020} and \citet{lacour2020} to place constraints on the presence of exomoons and binary planets in the \(\beta\) Pictoris system using high-precision astrometric measurements from the GRAVITY instrument. While we find no evidence supporting the existence of an exomoon, our analysis establishes strong constraints on the mass and orbital period of any potential companions, providing a framework for applying astrometric techniques to future exomoon searches. Our paper is organized as follows: Section \ref{observationsanalysis} outlines the instrumentation and data. Section \ref{analysis} describes our analysis methods, including the development and application of our model. Section \ref{results} presents our results, focusing on the derived mass and orbital constraints. Section \ref{discussion} then discusses the implications of our findings, comparing astrometry with complementary techniques and addressing the potential for future studies. We conclude in Section \ref{Conclusions}.

\section{Observations}\label{observationsanalysis}
%\subsection{Instrumentation}\label{instrumentation}
In this work, we used existing archival direct imaging and interferometric observations of \(\beta\) Pictoris b and c. \(\beta\) Pictoris has been observed for over a decade by numerous instruments, and the precision of astrometric measurements has increased over time. We focused on data taken with the GRAVITY interferometer \citep{gravity2017} and the SPHERE high-contrast imager \citep{beuzit2019}, provided in Table \ref{table:astrometric-measurements}. Following \bedit{\citet{lacour2021}}, we excluded earlier, less-precise data from the NACO (NAOS-CONICA) instrument \citep{lenzen2003, rousset2003}. Together, these datasets provide a high-precision foundation for our orbital analysis.

%, which significantly improved our residual plots.  was performed using instruments such as GRAVITY and SPHERE. , and NACO. However, we excluded astrometric observations of \(\beta\) Pictoris b from the NACO (NAOS-CONICA) instrument \citep{lenzen2003, rousset2003}, which provides enhanced contrast adaptive optics (AO) imaging on the Very Large Telescope (VLT). During initial tests of our two-planet and moon model, we found that including NACO data reduced precision. 

\begin{table*}[!t]
\centering
\caption{Astrometric Measurements of Beta Pic b and c}
\label{table:astrometric-measurements}
\scriptsize
\begin{tabular}{cccccccc}
\hline
Planet Name & Epoch (MJD) & RA (mas) & RA Err (mas) & DEC (mas) & DEC Err (mas) & Instrument & Correlation \\
\hline
Beta Pic b & 56999.23 & -188.44 & 4.4 & -294.74 & 3.99 & SPHERE & 0 \\
Beta Pic b & 57058.09 & -178.94 & 2.85 & -279.85 & 2.81 & SPHERE & 0 \\
Beta Pic b & 57058.09 & -180.52 & 1.17 & -282.71 & 1.17 & SPHERE & 0 \\
Beta Pic b & 57296.34 & -143.09 & 3.14 & -219.89 & 2.98 & SPHERE & 0 \\
Beta Pic b & 57296.34 & -144.9 & 1 & -217.92 & 1 & SPHERE & 0 \\
Beta Pic b & 57356.25 & -132.93 & 4.23 & -202.45 & 3.98 & SPHERE & 0 \\
Beta Pic b & 57356.25 & -134.32 & 1.52 & -202.08 & 1.52 & SPHERE & 0 \\
Beta Pic b & 57382.18 & -128.07 & 1.75 & -196.38 & 1.75 & SPHERE & 0 \\
Beta Pic b & 57382.18 & -130.96 & 3.35 & -195.71 & 3.31 & SPHERE & 0 \\
Beta Pic b & 57407.12 & -124.33 & 2.66 & -190.28 & 2.56 & SPHERE & 0 \\
Beta Pic b & 57407.12 & -125.1 & 1.73 & -187.57 & 1.73 & SPHERE & 0 \\
Beta Pic b & 57474.02 & -112.57 & 1.48 & -168.41 & 1.48 & SPHERE & 0 \\
Beta Pic b & 57474.02 & -113.69 & 2.62 & -168.9 & 2.55 & SPHERE & 0 \\
Beta Pic b & 57494 & -110.39 & 3.79 & -164.8 & 3.57 & SPHERE & 0 \\
Beta Pic b & 57494 & -106.79 & 1.7 & -161.71 & 1.7 & SPHERE & 0 \\
Beta Pic b & 57647.38 & -83.45 & 2.11 & -116.3 & 2.11 & SPHERE & 0 \\
Beta Pic b & 57647.38 & -79.507 & 3.81 & -116.28 & 3.76 & SPHERE & 0 \\
Beta Pic b & 57675.35 & -77.605 & 3.56 & -108.2 & 3.51 & SPHERE & 0 \\
Beta Pic b & 57675.36 & -76.58 & 2.54 & -111.22 & 2.54 & SPHERE & 0 \\
Beta Pic b & 57710.25 & -63.92 & 4.73 & -83.45 & 4.73 & SPHERE & 0 \\
Beta Pic b & 57710.27 & -77.255 & 10.7 & -105.18 & 10.29 & SPHERE & 0 \\
Beta Pic b & 58378.2 & 69.87 & 5.08 & 123.143 & 5 & SPHERE & 0 \\
Beta Pic b & 58378.35 & 68.54 & 2.92 & 127.13 & 2.92 & SPHERE & 0 \\
Beta Pic b & 58383.378 & 68.47 & 0.05 & 126.38 & 0.07 & GRAVITY & -0.86 \\
Beta Pic b & 58409.31 & 74.61 & 1.64 & 135.05 & 1.64 & SPHERE & 0 \\
Beta Pic b & 58409.38 & 77 & 3.57 & 138.44 & 4.04 & SPHERE & 0 \\
Beta Pic b & 58467.19 & 85.57 & 1.22 & 152.36 & 1.22 & SPHERE & 0 \\
Beta Pic b & 58467.2 & 85.93 & 4.12 & 154.926 & 3.75 & SPHERE & 0 \\
Beta Pic b & 58552.99 & 105.93 & 1.59 & 183.03 & 1.59 & SPHERE & 0 \\
Beta Pic b & 58790.3 & 142.55 & 2.16 & 248.71 & 2.27 & SPHERE & 0 \\
Beta Pic b & 58790.3 & 143.83 & 1.07 & 246.24 & 1.07 & SPHERE & 0 \\
Beta Pic b & 58796.17 & 145.514 & 0.108 & 248.589 & 0.045 & GRAVITY & -0.85 \\
Beta Pic b & 58798.356 & 145.65 & 0.03 & 249.21 & 0.09 & GRAVITY & -0.44 \\
Beta Pic b & 58839.19 & 153.56 & 2.42 & 258.71 & 2.57 & SPHERE & 0 \\
Beta Pic b & 58839.2 & 151.25 & 1.17 & 259.87 & 1.17 & SPHERE & 0 \\
Beta Pic b & 58855.065 & 155.41 & 0.15 & 264.33 & 0.29 & GRAVITY & -0.71 \\
Beta Pic b & 58887.07 & 159.91 & 0.94 & 270.94 & 0.94 & SPHERE & 0 \\
Beta Pic b & 58887.07 & 161 & 1.83 & 272.59 & 1.92 & SPHERE & 0 \\
Beta Pic b & 58889.139 & 160.96 & 0.06 & 273.41 & 0.13 & GRAVITY & -0.56 \\
Beta Pic b & 59221.238 & 211.59 & 0.02 & 352.62 & 0.05 & GRAVITY & -0.1 \\
Beta Pic b & 59453.395 & 240.63 & 0.09 & 397.89 & 0.04 & GRAVITY & -0.91 \\
\hline
Beta Pic c & 58889.14 & -67.36 & 0.17 & -112.59 & 0.24 & GRAVITY & -0.8 \\
Beta Pic c & 58891.065 & -67.67 & 0.11 & -113.2 & 0.19 & GRAVITY & -0.54 \\
Beta Pic c & 58916.043 & -71.88 & 0.07 & -119.6 & 0.14 & GRAVITY & -0.52 \\
Beta Pic c & 59220.163 & -52 & 0.2 & -80.86 & 0.34 & GRAVITY & -0.26 \\
\hline
\end{tabular}
\end{table*}

\subsection{SPHERE}\label{sphere}
Observations of \(\beta\) Pictoris b were acquired with the Spectro-Polarimetric High-contrast Exoplanet Research (SPHERE) instrument \citep{beuzit2019}. SPHERE is equipped with the Infra-Red Dual-band Imager and Spectrograph \citep[IRDIS,][]{Dohlen2008SPIE} and the Integral Field Spectrograph \citep[IFS,][]{Claudi2008}, enabling simultaneous high-contrast imaging. We make use of 34 SPHERE astrometric measurements from three different programs.  

23 SPHERE measurements were obtained between December 2014 and September 2018 as part of the SHINE Guaranteed Time Observations survey and are reported in \citet{Lagrange_2019_Post_conj}. These observations provided the pre-conjunction orbital coverage of the planet and used the IRDIFS mode, with IRDIS operating in the H2 (central wavelength = 1.59 $\mu$m; width = 0.055 $\mu$m) and H3 (central wavelength = 1.67 $\mu$m; width = 0.056 $\mu$m) narrow bands, except in December 2014 when the K1 (central wavelength = 2.1025 $\mu$m; width = 0.102 $\mu$m) and K2 (central wavelength = 2.255 $\mu$m; width = 0.109 $\mu$m) filters were used. An additional 5 measurements were obtained after conjunction as part of the SHINE follow-up program and are reported in \citet{lagrange2020}. These observations were taken in October 2018 and December 2018 using the IRDIFS configuration (IRDIS H2/H3 + IFS in YJ mode, 0.95–1.35 $\mu$m, $R\approx 54$), and March 2019 using the IRDIS K1/K2 filters with the IFS in YH mode (0.95–1.65 $\mu$m, $R\approx 33$). For the March 2019 epoch, only IFS yielded a usable measurement due to IRDIS saturation. The final 6 observations of the system were obtained under a PI program led by \citet{lagrange2020}, all using the H2/H3 + YJ setup as in the SHINE observations. These measurements were taken in November 2019, December 2019, and February 2020.

All SPHERE observations followed established reduction techniques outlined in \citet{chauvin2017} and \citet{delorme}. Specifically, the speckle calibration tool (SpeCal) developed for the SPHERE instrument by \citet{galicher2018} was used in the reduction process to retrieve the position and brightness information of the identified point sources.

\subsection{GRAVITY}\label{gravity}
The most important observations for our analysis (with the greatest ability to constrain exomoons) were taken with the GRAVITY instrument \citep{gravity2017} on the Very Large Telescope Interferometer (VLTI) at ESO's Paranal Observatory in Chile. GRAVITY collects K-band spectro-interferometric data and can achieve precise astrometric measurements with uncertainties ranging from 20$-$150 $\mu$as for directly imaged planets.

We use GRAVITY K band observations of \(\beta\) Pictoris b and c \bedit{originally obtained by \citet{lagrange2020} and \citet{lacour2021}, adopting the reprocessed astrometry presented in \citet{lacour2021} for all GRAVITY epochs}. The observations collected by \citet{lagrange2020} took place on six nights between September 22, 2018, and February 9, 2020. Three of the observations were obtained in the medium-resolution (R = 500) mode of the GRAVITY spectrometer, and two were obtained in the high-resolution mode (R = 4000). On January 7th, both modes were used. Following \citet{lacour2021}, we exclude two of the GRAVITY observations of \citet{lagrange2020} (MJDs 58738.267 and 58855.198) from our analysis due to poor coherence time and limited exposures.  We also include in our analysis additional observations collected by \citet{lacour2020} as part of the ExoGRAVITY program \citep{exogravity} on the night of January 6, 2021 for \(\beta\) Pictoris c and the nights of January 7 and August 27, 2021, for \(\beta\) Pictoris b. The astrometric calibration from both \citet{lagrange2020} and \citet{lacour2021} underwent processing using version 1.5.0 of the ESO GRAVITY pipeline \citep{lapeyrere2014}.

%due to preliminary residual plots indicating significant discrepancies compared to the data used by \citet{lacour2020}. 

%\newpage

\section{Analysis}\label{analysis}

\begin{table*}[!t]
    \centering
    \caption{Fit parameters and priors for both the ``Two Planet'' model and the ``Two Planet \& Moon'' model}
    \label{table:beta-pictoris-priors}
    \scriptsize
    \begin{tabular}{lcc}
        \hline
        Parameter & Two Planet Model & Two Planet \& Moon Model \\
        \hline
        $m_*$ (M$_{\odot}$) & $\mathcal{U}(1.19, 1.97)$ & $\mathcal{U}(1.19, 1.97)$ \\
        Parallax (mas) & $\mathcal{U}(48.76, 54.10)\times \mathcal{N}(51.44, 0.12)$ & $\mathcal{U}(48.76, 54.10)\times \mathcal{N}(51.44, 0.12)$ \\
        Error Scale Factor & $\mathcal{U}(0, 10)$ & $\mathcal{U}(0, 10)$ \\
        SPHERE Offset RA (mas) & $\mathcal{U}(-3, 3)$ & $\mathcal{U}(-3, 3)$ \\
        SPHERE Offset DEC (mas) & $\mathcal{U}(-3, 3)$ & $\mathcal{U}(-3, 3)$ \\
        \hline
        \multicolumn{3}{c}{Orbital Parameters for Beta Pic b} \\
        \hline
        $m_b$ (M$_{\text{Jup}}$) & $\mathcal{U}(5, 15)\times   \mathcal{N}(10.02, 0.29)$ & $\mathcal{U}(5, 15)\times   \mathcal{N}(10.02, 0.29)$ \\ 
        Semi-Major Axis (au) & $\mathcal{U}(9.63, 11.22)$ & $\mathcal{U}(9.63, 11.22)$ \\
        Eccentricity & $\mathcal{U}(0, 0.296)$ & $\mathcal{U}(0, 0.296)$ \\
        cos(Inclination) & $\mathcal{U}(0.0126, 0.0236)$ & $\mathcal{U}(0.0126, 0.0236)$ \\
        Argument of Periastron (deg) & $\mathcal{U}(181.0, 283.7)$ & $\mathcal{U}(181.0, 283.7)$ \\
        Longitude of Ascending Node (deg) & $\mathcal{U}(31.43, 32.12)$ & $\mathcal{U}(31.43, 32.12)$ \\
        Epoch of Periastron Passage & $\mathcal{U}(0.695, 0.973)$ & $\mathcal{U}(0.695, 0.973)$ \\
        \hline
        \multicolumn{3}{c}{Orbital Parameters for Beta Pic c} \\
        \hline
        $m_c$ (M$_{\text{Jup}}$) & $\mathcal{U}(5, 15)\times\mathcal{N}(9.03, 0.08)$ & $\mathcal{U}(5, 15)\times\mathcal{N}(9.03, 0.08)$ \\
        Semi-Major Axis (au) & $\mathcal{U}(1.66, 3.03)$ & $\mathcal{U}(1.66, 3.03)$ \\
        Eccentricity & $\mathcal{U}(0, 0.65)\times\mathcal{N}(0.231, 0.012)$ & $\mathcal{U}(0, 0.65)\times\mathcal{N}(0.231, 0.012)$ \\
        cos(Inclination) & $\mathcal{U}(-0.0407, 0.0843)$ & $\mathcal{U}({-}0.0407, 0.0843)$ \\
        Argument of Periastron (deg) & $\mathcal{U}(0, 102.4)$ & $\mathcal{U}(0, 102.4)$ \\
        Longitude of Ascending Node (deg) & $\mathcal{U}(29.55, 33.39)$ & $\mathcal{U}(29.55, 33.39)$ \\
        Epoch of Periastron Passage & $\mathcal{U}(0.417, 0.916)$ & $\mathcal{U}(0.417, 0.916)$ \\
        \hline
        \multicolumn{3}{c}{Orbital Parameters for Beta Pic b Moon} \\
        \hline
        $m_{\text{moon}}$ (M$_{\text{Jup}}$) & -- & $\mathcal{U}(0, 10)$ \\ 
        Period (days) & -- & $\mathcal{U}(0, 1500)$ \\
        cos(Inclination) & -- & $\mathcal{U}(-1, 1)$ \\
        Longitude of Ascending Node (deg) & -- & $\mathcal{U}(0, 360)$ \\
        Epoch of Periastron Passage & -- & $\mathcal{U}(0, 1)$ \\
        \hline
    \end{tabular}
\end{table*}

\subsection{Brightness Ratio}

We first investigated whether we must account for light emitted by any moon in our astrometric analysis. When we measure the astrometric wobble for two unresolved luminous sources, the motion of the ``photocenter,'' the center of flux of the two sources, will be smaller than the motion of either source. If a moon contributes significant flux compared to its host planet and we did not account for that in our analysis, we would infer an incorrect mass measurement or upper limit for that moon.   

\begin{figure} %  figure placement: here, top, bottom, or page
   \centering
   \includegraphics[width=\columnwidth]{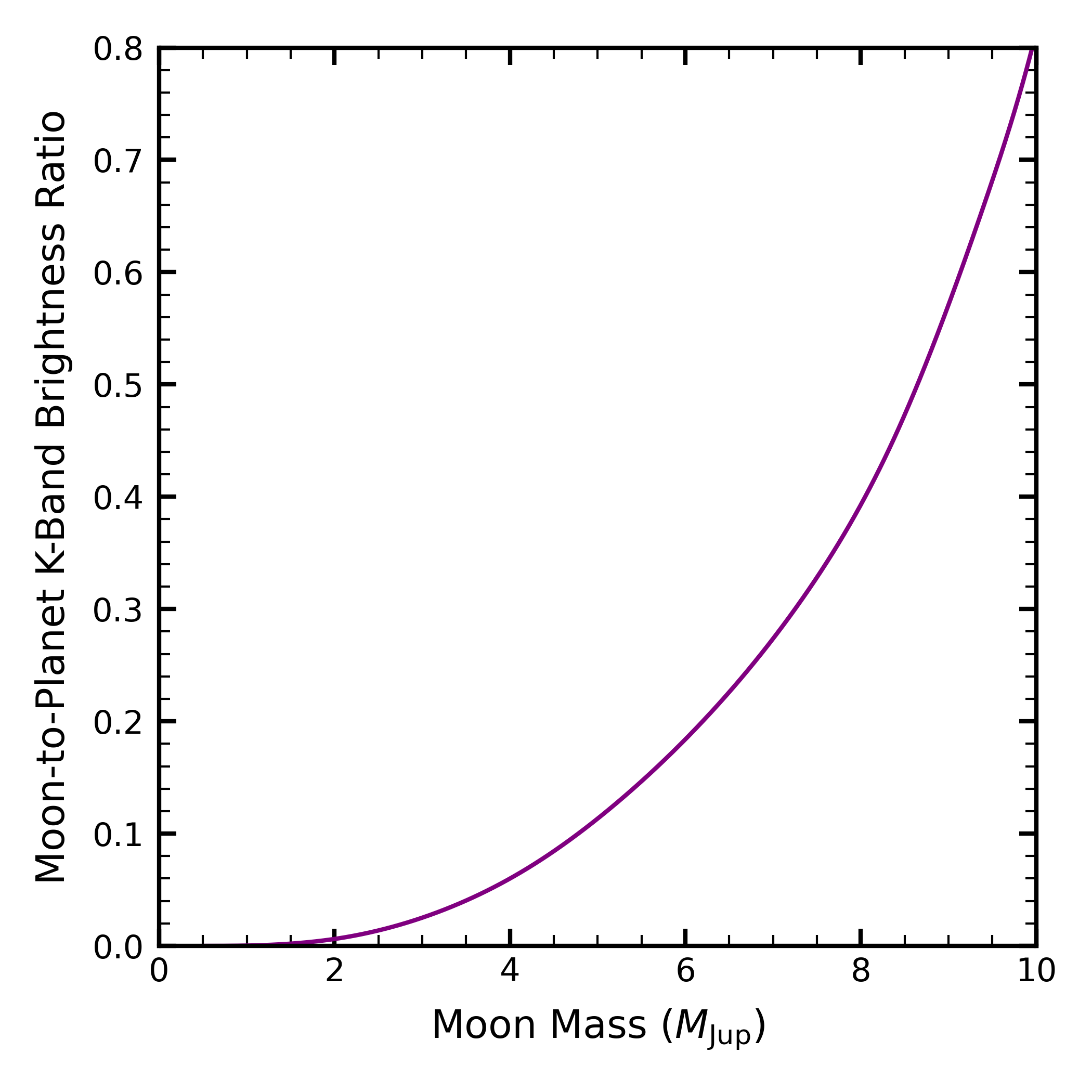} 
   \caption{Relationship between the K-band brightness ratio of \(\beta\) Pic b and a hypothetical moon as a function of the moon's mass. At lower moon masses, the K-band brightness contribution of the hypothetical moon is minimal compared to its host planet, motivating our assumption that any moons in the system are dark compared to the host planet.}
   \label{moon-brightness}
\end{figure}

%The astrometric signal measured by instruments such as GRAVITY and SPHERE corresponds to the motion of the photocenter, not the center of mass (COM). If a moon were bright relative to its host planet, it would shift the photocenter and reduce the observed astrometric signal. 
To evaluate the importance of this effect, we used the predicted K-band brightness of gas giant planets from ATMO evolutionary models \citep{Evolution_models}. For the moons, we took the K-band brightness of 23 Myr old gas giant planets with masses ranging from 0.5 $M_{\text{Jup}}$ to 10 $M_{\text{Jup}}$. We then compared these values to the predicted K-band brightness of a 10 $M_{\text{Jup}}$ planet at that age, analogous to \(\beta\) Pic b. The resulting brightness ratios are shown in Figure \ref{moon-brightness} as a function of the moon mass. Even for a 2 M$_{\text{Jup}}$ moon, we predict a brightness ratio $<5\%$. This corresponds to a moon-planet mass ratio of $\sim20\%$, larger than any found within our own solar system. We therefore predict that any moons around \(\beta\) Pic b will contribute negligible brightness, and that the photocenter's motion will track the position of the planet. We proceed under the assumption that any exomoon is effectively dark, minimizing its impact on the measured astrometric motion.

\subsection{Two Planet Model and Likelihood Function}
\label{sec: two-planet-model}
%\subsubsection{Model}
To fit the SPHERE and GRAVITY data, we constructed a model capable of accurately predicting the positions of \(\beta\) Pic b and c over time. To create this initial two-planet model, we adapted code from the \texttt{Orbitize!} Python package \bedit{\citep{Blunt2020AJ, blunt2024orbitizev3orbitfitting}} to simultaneously calculate the right ascension (RA) and declination (DEC) offsets of both planets with respect to the star. The model also accounts for the astrometric perturbation on \(\beta\) Pic b caused by its inner companion, \(\beta\) Pic c, following the \bedit{framework of \citet{lacour2021}}.

%ra_perturb1  = (m_c / (mtot2)) * raoff2_t1
%dec_perturb1 = (m_c / (mtot2)) * decoff2_t1

%Therefore, the model calculates both the displacement of \(\beta\) Pic A from its center of mass with \(\beta\) Pic c and incorporates this displacement into the position of \(\beta\) Pic b. This refinement incorporates the first-order planet-planet orbital effects, enhancing the model's precision.

We constructed a log likelihood function to compare the archival data (see Section \ref{observationsanalysis}) with this model. Maximizing the log likelihood (\(\ln{\mathcal{L}}\)) determines the best-fit orbital parameters for \(\beta\) Pic b and c, and exploring the parameter space constrains the plausible parameter space for any orbiting exomoon. \bedit{We allowed for the possibility that the uncertainties on the GRAVITY observations may be underestimated by incorporating a multiplicative ``jitter'' term, enabling the fit} to adjust the GRAVITY uncertainties such that: 

\begin{equation}\label{jitters}
\begin{array}{l}\sigma_{RA,b, gravity} = j \sigma_{RA,b,gravity,reported} \\
\sigma_{RA,c,gravity} = j \sigma_{RA,c, gravity, reported} \\
\sigma_{DEC,b,gravity} = j \sigma_{DEC,b, gravity, reported} \\
\sigma_{DEC,c,gravity} = j \sigma_{DEC,c, gravity, reported} \\
\end{array}
\end{equation}

\noindent where $j$ is the multiplicative jitter factor, and $\sigma$ represents uncertainties in either RA, or DEC for either planet b (\(\beta\) Pic b) or planet c (\(\beta\) Pic c). %our likelihood function to allow for To ensure the probabilities in our model are valid and accurately scaled, we performed a likelihood normalization calculation. Additionally, we incorporated a jitter term to address underestimated uncertainties, including astrophysical and instrumental contributions. This jitter term scaled the RA and DEC uncertainties for \(\beta\) Pic b,  improving the likelihood calculation, particularly for GRAVITY instrument observations.

The log likelihood consists of three components: the contributions from observations of \(\beta\) Pic b and c ($\ln\mathcal{L}_b$ and $\ln\mathcal{L}_c$), and Gaussian priors on various parameters, including the parallax and the masses and eccentricities of the planets. The final form of \(\ln{\mathcal{L}}\) is expressed as:

%\vspace{-15mm}
\begin{equation}
    \ln{\mathcal{L}} = \left(\ln\mathcal{L}_b + \ln\mathcal{L}_c \right) + \text{(priors)}
\end{equation}
%\vspace{-15mm}

\noindent where $\ln\mathcal{L}_b$ and $\ln\mathcal{L}_c$ are defined as:

\setlength{\jot}{10pt} 
\begin{equation}\label{chisq1}
\begin{split}
    \ln\mathcal{L}_b &= -0.5 \sum_{i} \left( \frac{1}{1 - \rho_{b,i}^2} \right) \Bigg[ \frac{(\text{RA}_{b,i} - \text{RA}_{\textit{model,b,i}})^2}{\sigma_{\textit{RA,b,i}}^2} \\ 
    &\quad + \frac{(\text{DEC}_{b,i} - \text{DEC}_{\textit{model,b,i}})^2}{\sigma_{\textit{DEC,b,i}}^2} \\
    &\quad - \frac{2\rho_{b,i}(\text{RA}_{b,i} - \text{RA}_{\textit{model,b,i}})(\text{DEC}_{b,i} - \text{DEC}_{\textit{model,b,i}})} 
    {\sigma_{\textit{RA,b,i}}\sigma_{\textit{DEC,b,i}}} \\
    &\quad + 2\ln{\sigma_{\textit{RA,b,i}}\sigma_{\textit{DEC,b,i}}} + \ln{(1 - \rho_{b,i}^2)}\Bigg] \\   
\end{split}
\end{equation}
\setlength{\jot}{5pt}

\setlength{\jot}{10pt} \
\begin{equation}\label{chisq2}
\begin{split}
    \ln\mathcal{L}_c &= -0.5 \sum_i \left( \frac{1}{1 - \rho_{c,i}^2} \right) \Bigg[ \frac{(\text{RA}_{c,i} - \text{RA}_{\textit{model,c,i}})^2}{\sigma_{\textit{RA,c,i}}^2} \\ 
    &\quad + \frac{(\text{DEC}_{c,i} - \text{DEC}_{\textit{model,c,i}})^2}{\sigma_{\textit{DEC,c,i}}^2} \\
    &\quad - \frac{2\rho_{c,i}(\text{RA}_{c,i} - \text{RA}_{\textit{model,c,i}})(\text{DEC}_{c,i} - \text{DEC}_{\textit{model,c,i}})} 
    {\sigma_{\textit{RA,c,i}}\sigma_{\textit{DEC,c,i}}} \\
    &\quad + 2\ln{\sigma_{\textit{RA,c,i}}\,\sigma_{\textit{DEC,c,i}}} + \ln{(1 - \rho_{c,i}^2)}\Bigg] \\   
\end{split}
\end{equation}
\setlength{\jot}{5pt} 

%%% Moon Brightness

In both Equations \ref{chisq1} and \ref{chisq2}, the summations are taken over all observational astrometric epochs $i$. \(\text{RA}_b\) and \(\text{DEC}_b\) in Equation \ref{chisq1} refer to the observed astrometric offsets in right ascension and declination for \(\beta\) Pic b, while \(\text{RA}_c\) and \(\text{DEC}_c\) in Equation \ref{chisq2} refer to the same for \(\beta\) Pic c. \(\text{RA}_{\text{model, b}}\), \(\text{DEC}_{\text{model, b}}\), \(\text{RA}_{\text{model, c}}\), \(\text{DEC}_{\text{model, c}}\) correspond to the predicted positions of \(\beta\) Pic b and c. \(\rho_b\) and \(\rho_c\) are the correlation coefficients for the astrometric measurements, and \(\sigma_{{RA,b}}\), \(\sigma_{{RA,c}}\), \(\sigma_{{DEC,b}}\), and \(\sigma_{{DEC,c}}\) are the uncertainties in right ascension and declination for planet b and c . We note that the $\sigma$ parameters in these equations are scaled by the ``jitter'' term described in Equation \ref{jitters}.  We are not aware of any derivation of the normalized likelihood function for data points with correlated uncertainties \bedit{between RA and DEC} (allowing us to fit for the best scaling factor for the correlated GRAVITY uncertainties), so we perform this calculation ourselves and describe it in detail in Appendix~\ref{AppendixA}.  

\begin{figure}[htb] %  figure placement: here, top, bottom, or page
   \centering
   \includegraphics[width=\linewidth]{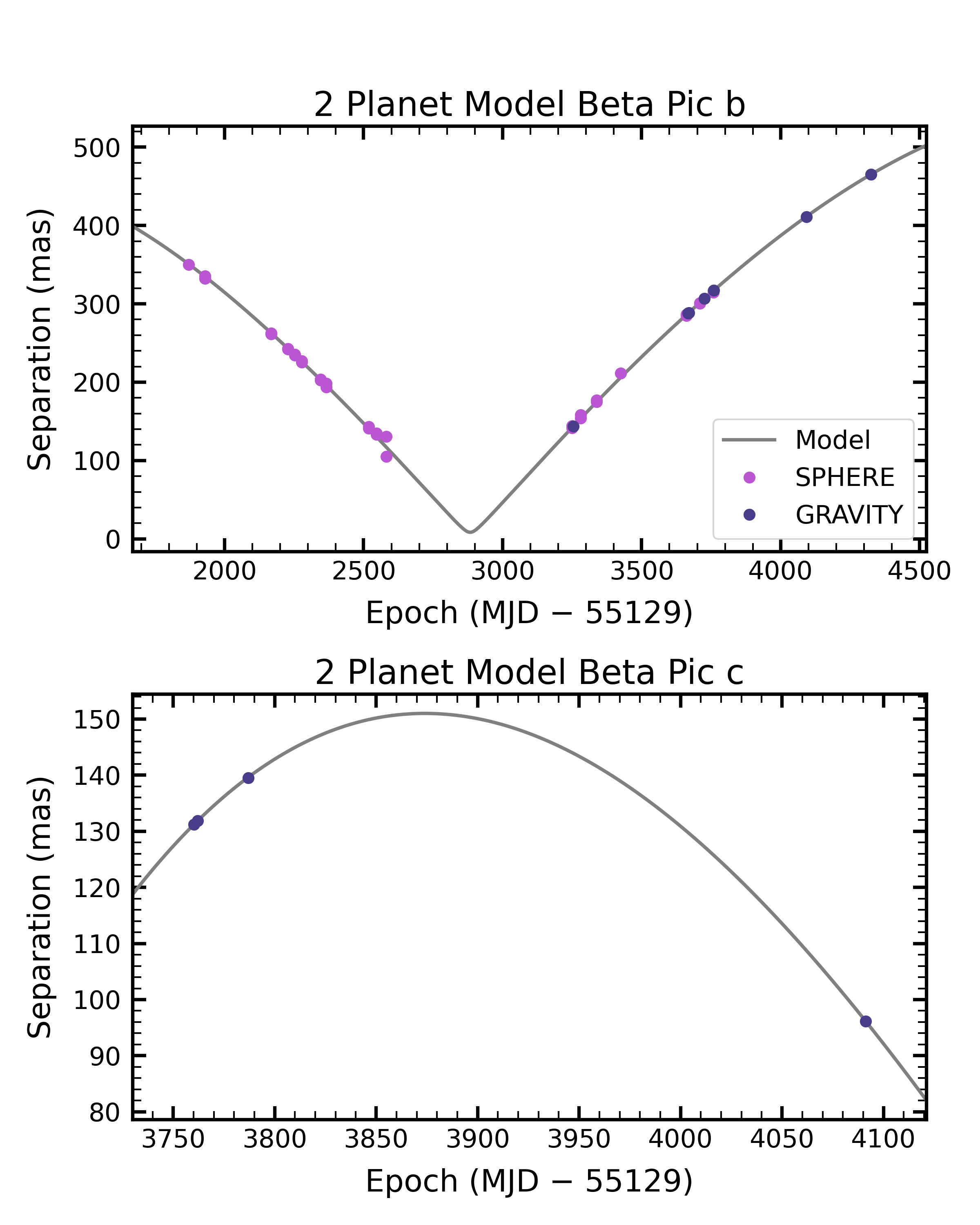} 
   \caption{The predicted separation of both \(\beta\) Pic b and c in our best-fit 2-planet model (solid grey curves), along with the actual observed separations. The dark purple points represent GRAVITY observations, while the lighter purple points represent SPHERE observations.}
   \label{BPB_2Planet_Sep}
\end{figure}

We imposed informative Gaussian priors based on pulsation-corrected RV data from \citet{2019_Lagrange} on the masses of $\beta$ Pic b ($0.00956 \pm 2.8 \times 10^{-4} \, M_\odot$) and c ($0.00862 \pm 7.6 \times 10^{-5} \, M_\odot$), as well as the eccentricity of $\beta$ Pic c ($0.231 \pm 0.012$). Additionally, we imposed the same Gaussian prior on the star's trigonometric parallax as \citet{lacour2021} ($51.44 \pm 0.12$ mas). For all other parameters, we used uniform priors with boundaries chosen either based on physical constraints (e.g., masses must be non-negative) or so that the posterior distributions for well-constrained parameters remain far from the boundaries. In total, our ``Two Planet'' model comprises a total of 19 free parameters. The full list of fitted parameters and priors we used is summarized in Table \ref{table:beta-pictoris-priors}, and the best fit model is compared to observations in Figure \ref{BPB_2Planet_Sep}.  % The total likelihood includes all these contributions, providing constraints on the orbital parameters of the \(\beta\) Pictoris system.

\subsection{Two Planet \& Moon Model} 

To model the astrometric signature of a hypothetical dark exomoon orbiting \(\beta\) Pic b, we extend the two-planet model described in Section \ref{sec: two-planet-model} by adding a Keplerian companion on a bound orbit around \(\beta\) Pic b. We used the same log likelihood functions as described in Equations \ref{jitters} through \ref{chisq2}, but include an additional astrometric perturbation term representing the motion of \(\beta\) Pic b about the barycenter of the \(\beta\) Pic b-moon system. To do this, we used \texttt{orbitize} to generate Keplerian position offsets ($\Delta$RA, $\Delta$DEC) for an exomoon orbiting \(\beta\) Pic b, where its gravitational influence shifts the apparent position of \(\beta\) Pic b by:

\begin{equation}
\begin{aligned}
\Delta \mathrm{RA}_{COM} = -\frac{m_{moon}\Delta \mathrm{RA}_{moon}}{m_b + m_{moon}} \\
\Delta \mathrm{DEC}_{COM} = -\frac{m_{moon}\Delta \mathrm{DEC}_{moon}}{m_b+m_{moon}}
\end{aligned}
\end{equation}

\noindent where $m_{moon}$ and $m_b$ are the moon and planet masses, and $\Delta\mathrm{RA_{COM}}$ and $\Delta\mathrm{DEC_{COM}}$ represent the astrometric offsets of the planet relative to the planet-moon center of mass (COM). These offsets are proportional to the moon's position offsets ($\Delta\mathrm{RA_{moon}, \Delta\mathrm{DEC_{moon}}}$) and scale by the ratio of the moon's mass to the planet's mass ($m_{moon}/m_b$). %The resulting combined astrometric motion of $\beta$ Pic b in the Two Planet \& Moon model is then expressed as:

%\begin{equation}
%\begin{aligned}
%\mathrm{RA_b} = \mathrm{RA_{b,0}} + \Delta \mathrm{RA_{perturb}} + \Delta \mathrm{RA_{COM}} \\
%\mathrm{DEC_b} = \mathrm{DEC_{b,0}} + \Delta \mathrm{DEC_{perturb}} + \Delta \mathrm{DEC_{COM}}
%\end{aligned}
%\end{equation}

%\noindent where $\mathrm{RA_{b,0}}$ and  $\mathrm{DEC_{b,0}}$ are the baseline right ascension and declination of $\beta$ Pic b corresponding to its unperturbed Keplerian orbit around the star. The terms $\Delta\mathrm{RA_{perturb}}$ and $\Delta\mathrm{DEC_{perturb}}$ represent the perturbation induced by $\beta$ Pic c, and the COM terms account for the astrometric wobble of $\beta$ Pic b caused by the gravitational influence of its orbiting moon.

We restricted our analysis to moons on circular orbits around \(\beta\) Pic b and confined the moon's orbital period to non-negative dynamically stable values based on the analysis of \citet{dynamical_stability_equation}. In particular, the upper limit was derived based on the stability of prograde satellites, which must remain within:

\begin{equation}
    R_{\text{max,p}} = R_{\text{Hill}} \times 0.4895 \left( 1 - 1.0305e_p - 0.2738e_m \right)
\end{equation}

\noindent where \(R_{\text{max,p}}\) is the maximum prograde stable orbital radius, \(R_{\text{Hill}}\) is the Hill sphere radius, \(e_p\) is the planet's orbital eccentricity, and \(e_m\) is the exomoon's orbital eccentricity (assumed to be zero). For \(\beta\) Pic b, the prograde stability limit corresponds to an orbital radius of 0.532 AU and an orbital period of 1451.64 days. Based on these calculations, we impose a uniform prior on the exomoon's orbital period with a lower bound defined by the Roche limit and an upper limit of 1500 days (which we truncated to the stability limit in post-processing). Overall, the addition of a potential exomoon requires five extra free parameters (moon mass, orbital period, inclination, position angle, and a reference time for the orbit), bringing the total number of fitted parameters from 19 in the two-planet model to 24 in the two-planet \& moon model (see Table \ref{table:beta-pictoris-priors}).

%%%Residual Plots
\begin{figure*}[htb] 
   \centering
   \includegraphics[width=\linewidth]{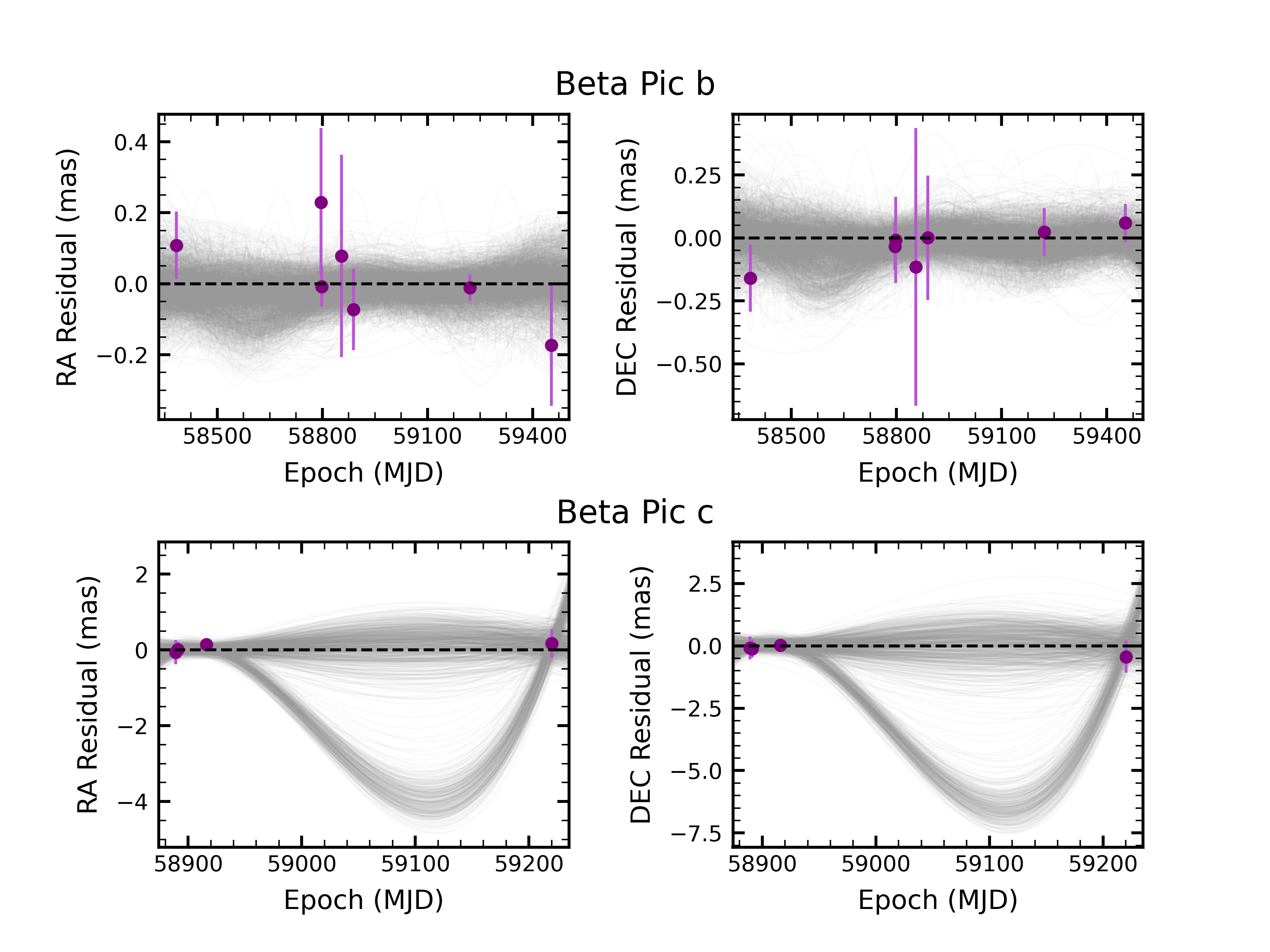} 
   \caption{Right ascension and declination residuals for the GRAVITY observations of $\beta$ Pictoris b and c with respect to our best-fit two-planet model. \bedit{Purple points and associated error bars mark the GRAVITY astrometric measurements, with uncertainties inflated by the best-fit error-scaling parameter.} The top row presents the residuals for $\beta$ Pictoris b, while the bottom row displays the corresponding residuals for $\beta$ Pictoris c. The gray lines represent 1000 randomly selected posterior draws from our ``Two Planet \& Moon'' fit. We see no correlated residuals to the ``Two Planet'' model indicative of an un-modeled signal, and we see no coherent pattern to the posterior draws for planet b, which includes hypothetical lunar perturbations. We note that the posterior samples of \betapicc\ show some evidence for bimodality due to the relatively sparse sampling of the planet's orbit with GRAVITY. }
   \label{residual_plots}
\end{figure*}

\begin{table*}[!t]
    \centering
    \caption{Best-fit parameters and uncertainties for the \betapictoris\ planets. We compare the ``Two Planet'' model, the ``Two Planet \& Moon'' model, and for reference, published results from \citet{lacour2021}.}
    \label{table:beta-pictoris-parameters-comparison}
    \scriptsize
    \begin{tabular}{llll}
        \hline
        Parameter & Two Planet System & Two Planet \& Moon System & \citet{lacour2021} \\
        \hline
        %$m_{\text{moon}}$ (M$_{\text{Jup}}$) & -- & \mmoonone\ummoonone & -- \\
        $m_b$ (M$_{\text{Jup}}$) & \mbnomoon\umbnomoon & \mb\umb & 10 \\ 
        $m_c$ (M$_{\text{Jup}}$) & \mcnomoon\umcnomoon & \mc\umc & 9.15$^{+1.08}_{-1.06}$ \\
        $m_*$ (M$_{\odot}$) & \mstarnomoon\umstarnomoon & \mstar\umstar & 1.73$^{+0.03}_{-0.02}$ \\
        Parallax (mas) & \plxnomoon\uplxnomoon & \plx\uplx & 51.44$^{+0.12}_{-0.12}$ \\
        Error Scale Factor & \errscalegravitynomoon\uerrscalegravitynomoon & \errscalegravity\uerrscalegravity & -- \\
        SPHERE Offset RA (mas) & \raoffsetspherebnomoon\uraoffsetspherebnomoon & \raoffsetsphereb\uraoffsetsphereb & -- \\
        SPHERE Offset DEC (mas) & \decoffsetspherebnomoon\udecoffsetspherebnomoon & \decoffsetsphereb\udecoffsetsphereb & -- \\
        \hline
        \multicolumn{4}{c}{Orbital Parameters for Beta Pic b} \\
        \hline
        Semi-Major Axis (au) & \smaonenomoon\usmaonenomoon & \smaone\usmaone & 9.95$^{+0.03}_{-0.02}$ \\
        Eccentricity & \ecconenomoon\uecconenomoon & \eccone\ueccone & 0.106$^{+0.004}_{-0.004}$ \\
        Inclination (deg) & \inconenomoon\uinconenomoon & \incone\uincone & 88.99$^{+0.01}_{-0.01}$ \\
        Argument of Periastron (deg) & \aoponenomoon\uaoponenomoon & \aopone\uaopone & 203.2$^{+2.8}_{-3.2}$ \\
        Longitude of Ascending Node (deg) & \panonenomoon\upanonenomoon & \panone\upanone & 31.80$^{+0}_{-0.01}$ \\
        Epoch of Periastron Passage & \tauonenomoon\utauonenomoon & \tauone\utauone & 0.730$^{+0.009}_{-0.010}$ \\
        \hline
        \multicolumn{4}{c}{Orbital Parameters for Beta Pic c} \\
        \hline
        Semi-Major Axis (au) & \smatwonomoon\usmatwonomoon & \smatwo\usmatwo & 2.61$^{+0.06}_{-0.06}$ \\
        Eccentricity & \ecctwonomoon\uecctwonomoon & \ecctwo\uecctwo & 0.32$^{+0.03}_{-0.02}$ \\
        Inclination (deg) & \inctwonomoon\uinctwonomoon & \inctwo\uinctwo & 88.92$^{+0.10}_{-0.10}$ \\
        Argument of Periastron (deg) & \aoptwonomoon\uaoptwonomoon & \aoptwo\uaoptwo & 60.8$^{+4.2}_{-3.7}$ \\
        Longitude of Ascending Node (deg) & \pantwonomoon\upantwonomoon & \pantwo\upantwo & 31.07$^{+0.05}_{-0.04}$ \\
        Epoch of Periastron Passage & \tautwonomoon\utautwonomoon & \tautwo\utautwo & 0.708$^{+0.012}_{-0.012}$ \\
        \hline
        %\multicolumn{4}{c}{Orbital Parameters for Beta Pic b Moon} \\
        %\hline
        %Period (days) & -- & \periodmoonone\uperiodmoonone & -- \\
        %Inclination (deg) & -- & \incmoonone\uincmoonone & -- \\
        %Argument of Periastron (deg) & -- & \panmoonone\upanmoonone & -- \\
        %Epoch of Periastron Passage & -- & \taumoonone\utaumoonone & -- \\
        %\hline
    \end{tabular}
\end{table*}

\begin{figure*}[htb]
   \centering
   \includegraphics[width=\linewidth]{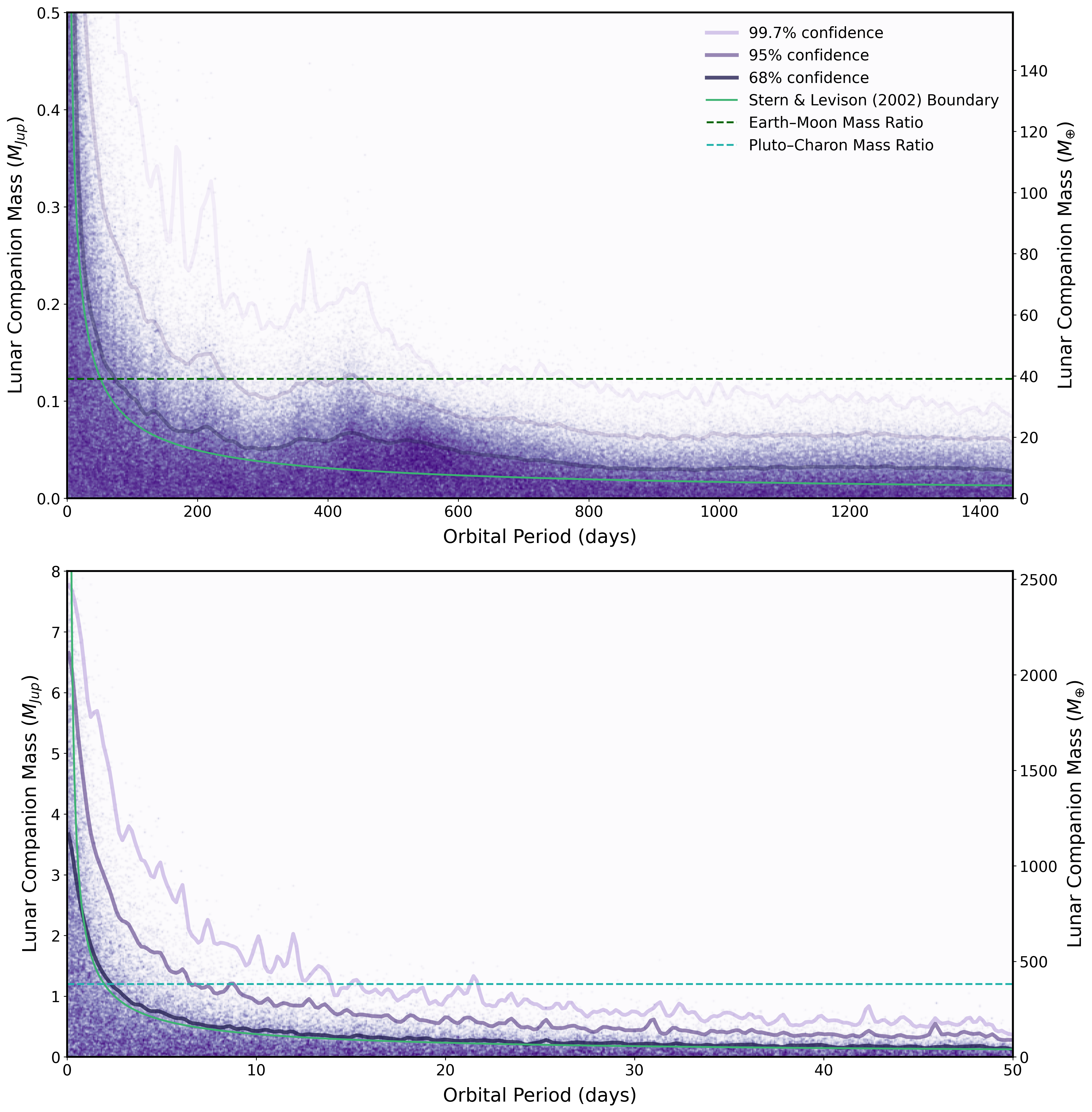} 
   \caption{Upper mass limits for a potential lunar companion to \(\beta\) Pic b as a function of orbital period. The color is proportional to the density of MCMC posterior samples, and the solid curves indicate the 68\%, 95\%, and 99.7\% confidence intervals. The \textbf{upper panel} covers longer orbital periods (\(\lesssim 1450\) days, within the dynamical stability limit), where companions more massive than \(\sim 0.5~M_{\text{Jup}}\) (\(\sim 60~M_{\oplus}\)) are excluded. The dashed line marks the Earth–Moon mass ratio scaled to \(\beta\) Pic b, and the solid green curve shows the moon/binary planet boundary from \citet{Stern&levison}, which marks the theoretical separation between large moons and binary planetary systems based on mass ratios and orbital stability criteria. The \textbf{lower panel} zooms in on short orbital periods (\(\lesssim 50\) days), where the constraints are weaker. The dashed line indicates the Pluto–Charon mass ratio, while the solid green curve again shows the \citet{Stern&levison} moon/binary planet boundary.}
   \label{purpleplot}
\end{figure*}

\subsection{Fitting}

We fitted the data with our astrometric models in three steps to determine both best-fit parameters and uncertainties. First, we performed an exploratory MCMC analysis on our ``Two Planet'' model using the Python implementation of the differential evolution MCMC sampler \citep[\texttt{edmcmc},][]{edmcmc}. This analysis successfully constrained the orbital parameters for the two-planet system, but was slow to achieve convergence due to the high-dimensional parameter space and the challenging covariances between different parameters. 

Second, we used the best-fit values and posterior distributions for the orbital parameters to determine the prior ranges for a fit with pocoMC \citep{karamanis2022accelerating, karamanis2022pocomc}, a Python implementation of a Preconditioned Monte Carlo algorithm. PocoMC uses machine learning to learn approximations of the posterior distributions that can substantially increase the efficiency of exploring parameter space. The result is a dramatic increase in the convergence speed for complex models like ours, particularly when incorporating the additional complexity of the exomoon parameters. We ran PocoMC on our ``Two Planet'' model using $n_{\mathrm{active}} = 600$ active particles and required at least $n_{\mathrm{effective}} = 2000$ effective particles per update step. Convergence was determined by monitoring the global effective sample size (ESS) and ensuring that it reaches a total of $n_{\mathrm{total}} = 480{,}500$ independent samples, indicating that the parameter chains are well-mixed and representative of the posterior distribution. We also examined the corner plot for each parameter, confirming sufficient sampling in the two-planet model. The model aligns well with the observed data (as shown in Figure \ref{BPB_2Planet_Sep}), and our two-planet posterior distributions closely match those reported by \citet{lacour2021}. A full list of the best-fit parameters is given in Table \ref{table:beta-pictoris-parameters-comparison}. 

Finally, we initialized a new PocoMC fit for our ``Two Planet \& Moon'' model to establish upper limits for the mass of a putative moon orbiting \(\beta\) Pic b. For this fit, we adopted the same configuration as in the ``Two Planet'' model, using $n_{\mathrm{active}} = 600$ active particles and $n_{\mathrm{effective}} = 2000$ effective samples. The final run consisted of a total of $n_{\mathrm{total}} = 468{,}000$ independent samples. The best-fit parameters for the ``Two Planet \& Moon'' model are also given in Table \ref{table:beta-pictoris-parameters-comparison}.  %The posterior probability distributions from the MCMC sampler are depicted in Figure \ref{purpleplot}, characterizing the mass of the hypothetical lunar companion. 

\section{Results}\label{results}

\subsection{Constraints on Exomoons Around \betapicb}

First, we inspected the results of our astrometric fits to assess whether there is any evidence for an exomoon orbiting \betapicb. When we compare the residuals of our best-fit ``Two Planet'' model to the results of our ``Two Planet \& Moon'' model, as shown in Figure \ref{residual_plots}, we find that there are no significant remaining signals in the residuals of our two-planet fit. We also find that the posterior probability distribution for the mass of a moon in the ``Two Planet \& Moon'' model is consistent with 0, indicating no significant signal of a moon in our dataset. We therefore proceed to investigate the upper limits on the mass of any putative exomoons orbiting \betapicb\ from our ``Two Planet \& Moon'' fits. 

We show the two-dimensional posterior distribution of moon mass and orbital period, marginalized over all possible planetary and exomoon orbits, in Figure \ref{purpleplot}. In this figure, the color represents the posterior probability density, and we show corresponding \(1\sigma\), \(2\sigma\), and \(3\sigma\) confidence limits as solid lines. We find that for short orbital periods ($\approx50$ days), our analysis places an upper limit on exomoon masses of approximately \(\approx 180~M_{\oplus}\) (\(0.6~M_{\text{Jup}}\)) with 95\% confidence. At intermediate periods near \(300\) days, we rule out companions exceeding \(\approx 65~M_{\oplus}\) (\(0.2~M_{\text{Jup}}\)). At longer orbital periods, we place the tightest constraints on the moon mass: we constrain the maximum allowed mass to $\approx 50~M_{\oplus}$ ($0.15~M_{\text{Jup}}$) at $700$ days and $\approx 30~M_{\oplus}$ ($0.1~M_{\text{Jup}}$) at $1,100$ days. At the \(1\sigma\) level, our constraints rule out exomoons more massive than \(\sim0.05~M_{\text{Jup}}\) (\(\sim16~M_{\oplus}\)) across most of the orbital period range. At some orbital periods around $\sim$300-500 days, we have lower sensitivity to exomoons, likely due to the relatively sparse observing cadence of the GRAVITY observations. Our $1\sigma$, $2\sigma$, and $3\sigma$ upper limits on the masses of companions to \betapicb\ are listed in Table \ref{tab: percentile_stub}.

\subsection{Orbital Parameters for \betapicb\ and c}

Our fits also allow us to investigate the sensitivity of the \betapic\ orbital parameters to different assumptions and noise models. The results of our fits are summarized in Table \ref{table:beta-pictoris-parameters-comparison}. In general, our orbital parameters for \betapic\ b and c are consistent with the values reported by \citet{lacour2021}. We also find good agreement between the orbital parameters determined by our ``Two Planet'' model and our ``Two Planet \& Moon'' model, indicating that the presence of an unknown moon around \betapicb\ would likely not bias our understanding of the planet's orbit.

\section{Discussion}\label{discussion}

\subsection{Constraints on Exomoons and Binary Planets in Context}

In this work, we derive the first astrometric limits on exomoons and binary planets around \betapicb. The limits that we achieve are driven by the ultra-precise astrometric measurements from the GRAVITY interferometer, which achieves 1-2 orders of magnitude better precision than traditional adaptive optics imaging. Our limits are strongest at long orbital periods, as expected due to the amplitude of astrometric perturbations scaling linearly with the semimajor axis of the companion. At the longest-period stable moon orbits, we can rule out moons more massive than about 30 Earth masses, or about twice the mass of Neptune. 

The moons that we could detect in the \betapic\ system are not like the moons in our own solar system. The Solar System's most massive moon is Ganymede, which has a mass 50 times smaller than that of Earth, and 1500 times smaller than the lowest mass moon we can rule out at 3$\sigma$ around \betapicb. However, the moons and binary planets that we could detect in the \betapic\ system have similar mass ratios to some of the Solar System moons. At long orbital periods, we could detect moons with masses of about 1\%, which is smaller than both the Pluto/Charon and Earth/Moon mass ratios. It is worth noting that both the Earth/Moon and Pluto/Charon systems are believed to have formed via collision and capture of lunar material \citep{Denton2025NatGe, Canup2001Natur}; any moons we could detect might likely have formed similarly in their systems. 

\subsection{Are we sensitive to moons or binary planets?}

Given that we are only sensitive \footnote{\bedit{Throughout this section, we roughly define sensitivity as the ability to exclude companions above the 99.7\% (3$\sigma$) posterior upper limits shown in Figure \ref{purpleplot}}.} to Neptune-mass companions and larger, it is worth asking: are the objects that we could detect moons, or should they instead be considered binary planets? The classification of a satellite as a moon or a binary planet remains an open question, particularly when considering systems beyond our solar system. A simple way to distinguish moons and binary planets would be to choose a mass ratio between the two orbiting objects above which an orbiting body is a binary planet, and below which it is a moon. Such a strategy has been applied to the IAU definition of a planet \citep{LecavelierdesEtangs2022NewAR}, which states that the mass ratio must be large enough that stable orbits can exist around the L4 and L5 Lagrange points:  $\approx 1/25$, or more precisely $2/(25 + \sqrt{621})$. If we were to, by extension, apply this definition to distinguish moons from binary planets, our analysis is sensitive to some exomoons around \betapicb\ at orbital periods longer than about 100 days. 

However, there is no official IAU definition for a moon, so instead, other criteria have been utilized. According to one commonly cited definition from \citet{Stern&levison}, a moon is a satellite with a mass and semi-major axis small enough to keep the center of mass of the moon-planet system within the radius of the primary planet. In contrast, a binary planet is defined as a system in which the companion is sufficiently massive for the center of mass to extend beyond the primary planet. By this definition, we are not sensitive to exomoons around \betapicb. We show the \citet{Stern&levison} boundary between moons and binary planets in Figure \ref{purpleplot} as a solid green line. This boundary falls slightly below our 1$\sigma$ contour at all orbital periods and scales in a similar manner with orbital period to our sensitivity limits. This similar scaling is expected, since the \citet{Stern&levison} definition of a moon and the amplitude of astrometric signals both depend on the center of mass motion of the planet. 

However, both of these definitions have drawbacks. For instance, although Charon formally meets the requirement for a binary planet from both of these two definitions, it is still commonly referred to as a moon of Pluto.  Moreover, under the \citet{Stern&levison} definition, the same object at different orbital separations may be either a moon or a binary planet. Given orbital evolution, this implies that an orbiting companion may be both a moon and a binary planet at different times in the system's history. For instance, while the Earth/Moon system currently meets the \citet{Stern&levison} criterion for a moon, after the Moon's semimajor axis has increased by about 40\% due to tidal evolution, it would by this definition formally become a binary planet. It is also plausible that because planet radii contract over time, a young planet/moon system may evolve into a binary system as the planet radius becomes small enough that it does not encompass the system's center of mass. %Ideally, a precise distinction between moons and binary planets will not have
Ultimately, a precise distinction between moons and binary planets will likely have to wait until astronomers have a better understanding of the broader population beyond the solar system. %The diversity of exoplanets have  Because we have yet to confirm and characterize moons beyond our Solar System, it remains an open question whether they resemble our Moon or differ substantially. Until an exomoon or binary planet is directly discovered and studied, establishing strict classification criteria will remain difficult, but our results suggest that any potential companion to $\beta$ Pic b will be best classified as a moon rather than a binary planet.

\begin{table}%[!t]
\centering
\label{tab: percentile_stub}
\scriptsize
\begin{tabular}{cccc}
\hline
Orbital Period (d) & 68\% (M$_{\text{Jup}}$) & 95\% (M$_{\text{Jup}}$) & 99.7\% (M$_{\text{Jup}}$) \\
\hline
2.424749  & 1.306464 & 3.461267 & 5.558522 \\
7.2742474 & 0.853945 & 2.157875 & 3.674100 \\
12.123745 & 0.485129 & 1.122373 & 2.067631 \\
16.973244 & 0.336653 & 0.734944 & 1.370374 \\
21.822742 & 0.274056 & 0.590851 & 1.096990 \\
\hline
\end{tabular}
\caption{Upper limits on the mass of exomoons and binary planets around \betapicb\ as a function of orbital period.  The full table is available in the online journal. }
\end{table}

\subsection{Prospects for Improved Exomoon Searches in the Future}

As discussed in the preceding subsections, exomoon detection is still in its early stages, and we are not yet sensitive to Solar System-like moons around \betapicb. It is therefore worth considering whether a pathway exists to detecting smaller moons around directly imaged planets using astrometry or other techniques. 

There is already good potential for improvement to astrometric moon detection with existing facilities by collecting additional data. For example, our dataset only includes 7 GRAVITY observations of \betapicb. Typical datasets used to search for planets have considerably more observations; for instance, \citet{Sahlmann2013A&A} reported a planet candidate with 25 astrometric observations, and often planet discoveries using the radial velocity method have hundreds or thousands of observations \citep[e.g.][]{Feng2017AJ}. If more observations with GRAVITY were collected with the same observing strategy, we could plausibly expect the sensitivity to exomoons to improve by a factor of a few. % roughly with $\sqrt{N}$, where $N$ is the number of observations collected, . 

Another opportunity for improving sensitivity with astrometry will come from the planned GRAVITY+ upgrade to the GRAVITY interferometer. This upgrade will significantly enhance astrometric precision, making it possible to routinely achieve $10$--$30~\mu$as level uncertainties \citep{Bourdarot2022}. Compared to our existing dataset, which has a median uncertainty of 70 $\mu$as and a mean uncertainty of 100 $\mu$as, this upgrade could give an order-of-magnitude improvement in sensitivity. Since we are currently sensitive to $\approx 30 M_\oplus$ moons, an order of magnitude improvement would give sensitivity to moons as low as a few Earth masses at long periods. % approximately 30 Earth masses Such an increase in precision could make it possible to detect moons with masses as low as a few times that of Earth.  

%current capabilities and will directly benefit exomoon searches by tightening mass limits and enabling detections at lower mass ratios.

It will also be possible to probe a larger parameter space by combining astrometric observations with other techniques, in particular radial velocities \citep[e.g.][]{Vanderburg2018AJ, Lazzoni_2022, Ruffio2023AJ}. Even though both radial velocities and astrometry probe the ``wobble'' of the planet due to the moon, the signal amplitude of an astrometric orbit scales proportionally to the moon's semimajor axis $a$, while the amplitude of a radial velocity orbit scales with $a^{-1/2}$. Therefore, radial velocities are much more sensitive to short-period moons, while astrometry is more sensitive to long-period moons. Recently, Kenworthy et al. \textit{in prep} reported the first constraints on exomoons around \betapicb\ using radial velocity observations, achieving 350 $m\,s^{-1}$ precision and demonstrating sensitivity to sub-Saturn mass moons at short orbital periods. Overall, combining these short-period radial velocity constraints with strong astrometric constraints on long-period moons is a promising path forward to searching for moons at all orbital separations around directly imaged planets. This will be especially important as the next generation of high-resolution spectrographs on 30m class telescopes could achieve radial velocity precisions of order 10-40 m s$^{-1}$, sufficient sensitivity to detect moons with mass ratios as low as ~$10^{-4}$ at short orbital periods \citep{Ruffio2023AJ}.

%\bedit{In addition, simulations by \citet{Ruffio2023AJ} demonstrate that RV monitoring of self-luminous, directly imaged planets is already sensitive to large satellites with mass ratios of order 1-4\% using current instrumentation. Their analysis further shows that the next generation of high-resolution spectrographs on 30m class telescopes could achieve radial velocity precisions of order 10-40 m s$^{-1}$, allowing for sensitivity to moons with mass ratios as low as ~$10^{-4}$ at short orbital periods.}

Both of these techniques could considerably improve as new telescopes are commissioned. \citet{Vanderburg2018AJ} pointed out the opportunity to make extremely precise radial velocity observations of directly imaged planets with 30-meter class telescopes, where the improved angular resolution will dramatically improve the signal-to-noise ratio of planetary spectra. This could improve typical radial velocity precision from hundreds of meters per second to few $m\,s^{-1}$  precision like that achieved on stars today. On longer timescales, improving the capabilities of optical long-baseline interferometry could also aid in exomoon detection. For example, adding a fifth Unit Telescope to the VLTI (with a longer baseline) could considerably improve the angular resolution and astrometric precision of a successor to GRAVITY.

\section{Conclusions}
\label{Conclusions}
In this work, we used astrometric measurements from the GRAVITY interferometer and SPHERE imaging to place constraints on the presence of exomoons around $\beta$ Pic b. Our conclusions can be summarized as follows: 
%In particular, we establish upper limits on the presence of exomoons and binary planets orbiting $\beta$ Pic b as follows

    \begin{itemize}[topsep=1pt,itemsep=1pt,parsep=1pt,partopsep=1pt]
        \item For short orbital periods ($\approx 50$ days), we exclude moons more massive than $\approx 180~M_{\oplus}$ ($0.6~M_{\text{Jup}}$) at the $3\sigma$ level.
        \item At intermediate periods near $300$ days, we rule out companions exceeding $\approx 65~M_{\oplus}$ ($0.2~M_{\text{Jup}}$) at the $3\sigma$ level.
        \item At long orbital periods, we constrain the maximum allowed mass to $\approx 50~M_{\oplus}$ ($0.15~M_{\text{Jup}}$) at $700$ days and $\approx 30~M_{\oplus}$ ($0.1~M_{\text{Jup}}$) at $1,100$ days (both at the $3\sigma$ level).
    \end{itemize}

%, and $\approx 15~M_{\oplus}$ ($0.05~M_{\text{Jup}}$) at $1,400$ days at the $3\sigma$ level.

These are the first astrometric limits on the presence of exomoons and binary planets in the \betapictoris\ system, and to our knowledge, along with \citet{Kral2025arXiv}, one of the first astrometric constraints on companions around any directly imaged planet. We find that existing observations are primarily sensitive to systems where the center of mass is outside of the planet itself, which is a commonly cited criterion for distinguishing binary planets from moons, \citep{Stern&levison}. However, these observations are sufficient to achieve sensitivity to moon/planet mass ratios smaller than that of the Earth and its moon, and increasing the number of GRAVITY observations could plausibly extend sensitivity to companions that are unambiguously moons. 

The derived astrometric limits highlight the power of astrometry in constraining exomoon orbital parameters, particularly when combined with interferometric measurements. Looking forward, synergy with future GRAVITY+ data and complementary RV measurements could further tighten constraints and enable confident detections of exomoons.

\acknowledgments
We extend our gratitude to Sarah Blunt for her assistance in adapting code from the \texttt{Orbitize!} Python package to place constraints on exomoons in the Beta Pictoris system, and to Ellen Price for helpful advice on optimizing PocoMC. We also thank Pope Leo XIV for blessing an early draft of this manuscript, after which the final major bugs in our analysis were quickly resolved. 
This research was funded by MIT's Summer Research Program, the MIT Dean of Science Fellowship (Isabella Macias), and the National Science Foundation Graduate Research Fellowship under Grant No. 1745302 (Sydney Jenkins). Andrew Vanderburg is supported as a Sloan Research Fellow. A portion of the simulations was performed on the MIT-PSFC partition of the Engaging cluster at the MGHPCC facility (www.mghpcc.org), funded by DoE grant number DE-FG02-91-ER54109. This publication acknowledges the use of data products from the GRAVITY instrument from the Very Large Telescope Interferometer (VLTI) at ESO's Paranal Observatory, Chile. This research has made use of the Astrophysics Data System, funded by NASA under Cooperative Agreement 80NSSC21M0056.

\section*{Data Availability}
The percentile data used to generate Figure \ref{purpleplot} are included in this paper's supplementary material. An excerpt of the table is shown in Table \ref{tab: percentile_stub}.

\vspace{3mm}

\facility{VLTI, VLT:Melipal}

\software{matplotlib \citep{plt}, pocoMC \citep{karamanis2022accelerating, karamanis2022pocomc}, 
          numpy \citep{np}, {edmcmc \citep{edmcmc}}
          }

%% get the citations to show in the compiled file do the following:https://www.overleaf.com/project/5d7a78ba38b36000015f7725

\clearpage

\appendix

\section{Scaling Factor and Normalization}
\label{AppendixA}
This appendix details the derivation of the form of the normalization of a likelihood function that includes correlations between observations in the RA and DEC directions. Including the normalization term in the likelihood function we use in our astrometric fits allows us to adjust the scale of the correlated GRAVITY uncertainties within the fit. 

\subsection{Likelihood function}

We define the likelihood contribution $\mathcal{L}_i$ associated with observation $i$ to be: 

\begin{equation}
    \mathcal{L}_i = N_i e^{-\chi_i^2/2}
\end{equation}

\noindent where $N$ is a normalization constant and $\chi_i^2$ is an generalization of the traditional $\chi^2$ statistic to include correlated uncertainties. In particular, we define $\chi_i^2$ as: 

\begin{equation}\label{chisq_eq}
    \chi_i^2 = \frac{1}{(1 - \rho_i^2)} \left( \frac{(\text{RA}_i - \text{RA}_{\text{model},i})^2}{\sigma_{{RA},i}^2} + \frac{(\text{DEC}_i - \text{DEC}_{\text{model},i})^2}{\sigma_{{DEC},i}^2} - \frac{2\rho_i (\text{RA}_i - \text{RA}_{\text{model},i})(\text{DEC}_i - \text{DEC}_{\text{model},i})}{\sigma_{{RA},i} \sigma_{{DEC},i}} \right)
\end{equation}

As in Section \ref{sec: two-planet-model},  we define $\rho_i$ to be the correlation coefficient between astrometric measurements in the RA direction and the DEC direction for observation $i$. Again, $\text{RA}_i$ and $\text{DEC}_i$ refer to our astrometric measurements in those directions, while $\text{RA}_{\text{model},i}$ and $\text{DEC}_{\text{model},i}$ refer to our model astrometric measurements in those directions. 

As usual, to calculate the likelihood $\mathcal{L}$ of all observations together, we take the products of their individual likelihood contributions: 

\begin{equation}
\mathcal{L} = \prod_i{\mathcal{L}_i}
\end{equation}

\subsection{Derivation of the Normalization constant $N$}

Because likelihood contributions are probabilities, they must integrate to 1 over all possible values of $\text{RA}_i$ and $\text{DEC}_i$: 
\begin{equation}
    \int_{-\infty}^{\infty} \int_{-\infty}^{\infty} \mathcal{L}_i \ d\,\text{RA}_i \ d\,\text{DEC}_i = \int_{-\infty}^{\infty} \int_{-\infty}^{\infty} N_i e^{-\chi_i^2/2}  d\,\text{RA}_i \ d\,\text{DEC}_i = 1
\end{equation}

We ensure this condition holds by evaluating the integral expression and solving for the normalization constant $N_i$. We start by defining several constants and new variables for simplicity. In particular, we define: 

%\begin{equation}
%    N_i = \frac{1}{\int_{-\infty}^{\infty} \int_{-\infty}^{\infty} e^{-\chi_i^2/2}  d\,\text{RA}_i \ d\,\text{DEC}_i}
%\end{equation}

%To evaluate this expression, 

\begin{equation}
    f_i \equiv \frac{1}{1 - \rho^2_i}, \quad \Delta r_i \equiv \text{RA}_i - \text{RA}_{{model}, i}, \quad \Delta d \equiv \text{DEC}_i - \text{DEC}_{{model}_i}
\end{equation}

These allow us to rewrite our expression for $\chi^2_i$ more compactly as:

\begin{equation}
    \chi^2_i = f_i \left( \frac{\Delta r_i^2}{\sigma_{RA,i}^2} + \frac{\Delta d_i^2}{\sigma_{DEC,i}^2} - \frac{2\rho_i \Delta r_i \Delta d_i}{\sigma_{RA,i} \sigma_{DEC,i}} \right)
\end{equation}

We then define three more constants: 

\begin{equation}
    k_{1,i} \equiv \frac{f_i}{2\sigma_{RA,i}^2}, \quad k_{2,i} \equiv \frac{f_i}{2\sigma_{DEC,i}^2}, \quad k_{3,i} \equiv \frac{\rho_if_i}{\sigma_{RA,i} \sigma_{DEC,i}}
\end{equation}

\noindent and substitute them into our expression for the likelihood contribution: 

\begin{equation}
    \mathcal{L}_i = N_i e^{-k_{1,i} \Delta r_i^2} e^{-k_{2,i} \Delta d_i^2} e^{k_{3,i} \Delta r_i \Delta d_i}
\end{equation}

Returning to the normalization condition, we write: 

\begin{equation}
    \int_{-\infty}^{\infty} \int_{-\infty}^{\infty} \mathcal{L} \ d\Delta r \ d\Delta d = \int_{-\infty}^{\infty} \int_{-\infty}^{\infty} N_i e^{-k_{1,i} \Delta r_i^2} e^{-k_{2,i} \Delta d_i^2} e^{k_{3,i} \Delta r_i \Delta d_i} \ d\Delta r_i \ d\Delta d_i =  1
\end{equation}

By evaluating the multivariate Gaussian integral, we obtain the condition:

\begin{comment}
\begin{equation}
    \frac{2 N_i \pi}{\sqrt{k_{2,i}} \sqrt{4 k_{1,i} - \frac{k_{3,i}^2}{k_{2,i}}}} = 1
    %\quad \text{(if Re} \left[ \frac{k_{3,i}^2}{k_{2,i}} \right] < 4 \text{Re} \left[ k_{1,i} \right] \text{)}
\end{equation}

or 
\end{comment}

\begin{equation}
    \frac{2 N_i \pi}{\sqrt{4 k_{1,i}k_{2,i} - k_{3,i}^2}} = 1
    %\quad \text{(if Re} \left[ \frac{k_{3,i}^2}{k_{2,i}} \right] < 4 \text{Re} \left[ k_{1,i} \right] \text{)}
\end{equation}

Solving for \( N_i \):
\begin{comment}
\begin{equation}
    N_i = \frac{\sqrt{k_{2,i}} \sqrt{4k_{1,i} - \frac{k_{3,i}^2}{k_{2,i}}}}{2\pi}
\end{equation}

or
\end{comment}

\begin{equation}
    N_i = \frac{\sqrt{4k_{1,i}k_{2,i} - k_{3,i}^2}}{2\pi}
\end{equation}

Substituting our original definitions for $f_i$, $k_{1,i}$, $k_{2,i}$, and $k_{3,i}$, we obtain: %\( f_i = \frac{1}{1 - \rho_i^2} \)

\begin{equation}
    N_i = \frac{1}{2\pi \sigma_{RA,i}\, \sigma_{DEC,i} \sqrt{1 - \rho_i^2}}
\end{equation}

Including the normalization in this form leads to the log likelihood function we use in our modeling (Equations \ref{chisq1} and \ref{chisq2}). 

\vspace{10pt}
%\begin{comment}
%\subsection{Jitter Term}

%    #I want to add jitter term only to ra and dec errors belonging to GRAVITY from planet b
%    GRAVITY_b = np.argwhere(instrument_b == 0)
%    scale_err_ra1[GRAVITY_b] = scale_err_ra1[GRAVITY_b] * errscale_Gravity
%    scale_err_dec1[GRAVITY_b] = scale_err_dec1[GRAVITY_b] * errscale_Gravity

%%    #jittering term is included to all planet c errors bc they're all GRAVITY
%    scale_err_ra2 = scale_err_ra2 * errscale_Gravity
%    scale_err_dec2 = scale_err_dec2 * errscale_Gravity

%\begin{equation}
%    \text{GRAVITY RA}_1 = \text{GRAVITY RA}_1 * jitter

%    \text{GRAVITY RA}_2 = \text{GRAVITY RA}_@ * jitter

%    \text{DEC}_{\text{model, b}}
    
%\end{equation}
%\end{comment}

\clearpage

\vspace{5mm}
\bibliography{refs}{}
\bibliographystyle{aasjournal}

%    \bibliography{Bibliography}

%\begin{figure*}[htb] %  figure placement: here, top, bottom, or page
%   \centering
%   \includegraphics[width=\linewidth]{lc-crop.pdf} 
   %\includegraphics[width=\linewidth]{lc.eps} 
%   \caption{\Kepler\ light curve phase-folded on the transits of b (\textit{left}) and c (\textit{right}). Grey points are individual \Kepler\ long-cadence exposures, purple points are averages in phase, and the red solid line is the best-fit transit model.}
%   \label{lc}
%\end{figure*}



\end{document}